\documentclass[aps,prl,twocolumn,superscriptaddress,
groupedaddress]{revtex4}
\usepackage[T1]{fontenc}
\usepackage[latin9]{inputenc}
\usepackage{verbatim}
\usepackage{textcomp}
\usepackage{amsthm}
\usepackage{graphicx}
\usepackage{amsfonts}
\usepackage{amssymb}
\usepackage{times,txfonts}
\usepackage{hyperref}
\usepackage{color}

\usepackage{CJK}

\begin{document}

\title{Magneto-nonlinear  Hall effect in time-reversal breaking system}

\author{Anwei Zhang}
\email{zhanganwei@shnu.edu.cn}
\affiliation{Department of Physics, Mathematics $\&$ Science College, Shanghai Normal University, No. 100 Guilin Road, Shanghai, 200234 China}
\author{Jun-Won Rhim}
\email{jwrhim@ajou.ac.kr}
\affiliation{Department of Physics, Ajou University, Suwon 16499, Korea}
\affiliation{Research Center for Novel Epitaxial Quantum Architectures, Department of Physics, Seoul National University, Seoul, 08826, Korea}

\begin{abstract}
Magneto-nonlinear Hall effect is known to be intrinsic and requires time-reversal symmetry. Here we show that a new type of magneto-nonlinear Hall effect can occur
in the time-reversal breaking materials within the second-order response to in-plane electric and vertical magnetic fields.
%
%
Such a Hall response is generated by the oscillation of the electromagnetic field and has a quantum origin arising from a geometric quantity associated with the Berry curvature and band velocity.
%
We demonstrate that the massive Dirac model of LaAlO3/LaNiO3/LaAlO3 quantum well can be used to detect this Hall effect. 
Our work widens the theory of the Hall effect in the time-reversal breaking system by proposing a new kind of nonlinear electromagnetic response.

\end{abstract}


\maketitle


\section{I. Introduction} Recently, the second-order nonlinear Hall effects~\cite{gao2014field,liu2021intrinsic,wang2021intrinsic,bharti2022high,bhalla2022resonant,sodemann2015quantum,morimoto2016semiclassical,facio2018strongly,you2018berry,zhang2018berry,ma2019observation,kang2019nonlinear,gao2020second,watanabe2021chiral,du2021quantum,du2021nonlinear,wang2024orbital}, i.e., the current is proportional to the square of the electric field: $j_a \propto E_b E_c$, have attracted broad interest, due to their application in revealing band geometric quantities, characterizing crystal symmetries and probing $\mathrm{N\acute{e}el}$ vector~\cite{shao2020nonlinear} and quantum critical point~\cite{rostami2020probing}.
The second-order Hall responses can be divided into two types. 
%
%
%
One is attributed to the intrinsic mechanism of the band structure. It is independent of the relaxation time $\tau$ and the oscillating frequency $\omega$ of the fields and requires time-reversal broken~\cite{gao2014field,liu2021intrinsic,wang2021intrinsic,bharti2022high,bhalla2022resonant}. 
%
Another comes from the Berry curvature dipole.
%
It is an extrinsic effect due to its
dependence on $\tau$ and $\omega$. Besides, it respects time-reversal symmetry ~\cite{sodemann2015quantum,morimoto2016semiclassical,facio2018strongly,you2018berry,zhang2018berry,ma2019observation,kang2019nonlinear,gao2020second,du2021quantum,du2021nonlinear}.

In addition to the previous mentioned nonlinear Hall effect, it was shown that there is a different type of nonlinear Hall effect, named magneto-nonlinear  Hall effect~\cite{gao2014field,wang2024orbital} , in which the current is proportional to  electric field and magnetic field, i.e., $j_a \propto E_b B_c$. It is an intrinsic effect and  occurs in materials with time-reversal symmetry.  A question arises naturally:  whether there is a corresponding magneto-nonlinear  Hall effect in the Hall device that sustains the time-reversal broken?

In this paper, we give a postive answer to this question. We use a full quantum method, i.e., the Matsubara formalism, to predict a novel Hall effect which is proportional to
the electric and magnetic fields.
%
Here, we let the electric and magnetic fields be polarized along the $y$- and $z$-directions, respectively.
The Hall current in the $x$-axis is given by
\begin{equation}
j_x(2\omega)=\frac{2ie^3 }{\hbar\omega} D E_y B_z,
\end{equation}
where $E_y$ and $B_z$ are the amplitudes of the electromagnetic fields.
%
Most importantly, we show that the electromagnetic coefficient is proportional to a geometric quantity $D$ associated with the Berry curvature and band velocity.
%
%
The  Hall current found in this work also depends on the oscillating frequency of the external fields. 
For finite current, our system should be time-reversal broken,
 since in the presence of time-reversal symmetry, the Berry curvature and band velocity are all odd functions of momentum, and  the current vanishes.
Such a  Hall current can be detected in a massive Dirac model of LaAlO$_3$/LaNiO$_3$/LaAlO$_3$ quantum well.
%
%
Our Letter widens the theory for the Hall effect in the time-reversal breaking system.



\smallskip
\section{II.  Model}
\begin{figure}
\centering
\includegraphics[width=0.45\textwidth]{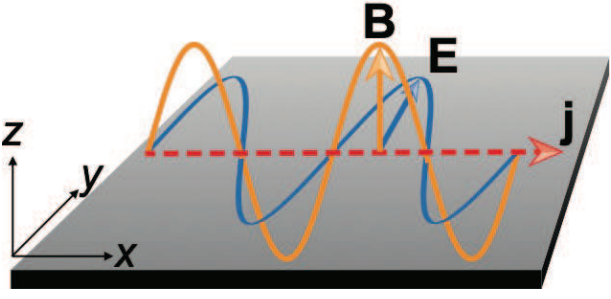}
\caption{Schematic of the system under consideration. The applied electric field $\mathbf{E}$ and magnetic field $\mathbf{B}$ are polarized in $y$- and $z$-direction, respectively. The induced Hall current $\mathrm{j}$ is perpendicular to the direction of the electromagnetic field. The fields propagate in the same direction as the Hall current.
}\label{figure.1}
\end{figure}

In this work, we consider a  clean system and apply an  electric field polarized in $y$-direction and propagating along $x$-direction, i.e., $\mathbf{E}=E_y\hat{y} e^{i q x-i \omega t}$, and a magnetic field polarized along $z$-direction, i.e.,
$\mathbf{B}=B_z\hat{z} e^{i q x-i \omega t}$, as illustrated in Fig.~\ref{figure.1}. 
%
Here, $q$ and $\omega$ are the moduli of the momentum and frequency of the fields, respectively.
Under the presence of such external fields, the system can be described by the vector potential
\begin{equation}\label{1}
\mathbf{A}=A_y\hat{y} e^{i q x-i \omega t},
\end{equation}
where $A_y$ satisfies the relations $A_y=E_y/i\omega =B_z/iq$.

The full Hamiltonian of the system in the presence of the external fields can be written as $H=H_0+H^{'}$. Here $H_0$ is the Hamiltonian of the sample and $H^{'}$ is the perturbed Hamiltonian which is given by Taylor series of the Hamiltonian in minimal coupling scheme \cite{morimoto2016semiclassical}
\begin{equation}\label{2}
    H^{'}=e\mathbf{\hat{v}}\cdot \mathbf{A}+\frac{e^2}{2}\partial_{{\mathbf{k}}}(\mathbf{\hat{v}}\cdot\mathbf{A}) \cdot \mathbf{A},
\end{equation}
where $-e$ is the charge of the electron and $\mathbf{\hat{v}}=\partial_{\mathbf{k}}H_0$ is the velocity operator. Since we consider second-order response in this paper, we expand the Hamiltonian up to second-order.
As a result, the current operator in our system will be described by
 \begin{equation}\label{2001}
 \hat{j}_x=-e\partial_{k_x}H=-e\hat{v}_x-e^2\partial_{k_x}(\mathbf{\hat{v}}\cdot \mathbf{A})-\frac{e^3}{2}\partial_{k_x}[\partial_{{\mathbf{k}}}(\mathbf{\hat{v}}\cdot\mathbf{A}) \cdot \mathbf{A}].
 \end{equation}
Note that we set $\hbar=1$ in the paper.

\section{III. Second-order nonlinear response}  
Now we consider the second-order harmonic response with respect to the vector potential.
By using second quantization and Fourier transformation, the expectation value of the  current $\langle\hat{j}_x(2\mathbf{q},2i\omega)\rangle$ in momentum-frequency space can be written as (see Appendix A)
\begin{equation}\label{3}
j_x(2\mathbf{q},2i\omega)=\frac{1}{\beta V}\sum_{\mathbf{k},i\omega_n}\mathrm{Tr}[\hat{j}_x G(\mathbf{k},i\omega_n;\mathbf{k}-2\mathbf{q},i\omega_n-2i\omega)],
\end{equation}
where $V$ denotes the volume of a three-dimensional system or the area of a two-dimensional system, $\beta=1/k_B T$ is the inverse temperature, $\mathbf{q}=q\hat{x}$, $\omega_n=(2n+1) \pi /\beta$ and
 $\omega=2m \pi /\beta$ are the fermionic and bosonic Matsubara frequencies, respectively.
According to Dyson equation, the exact
Green's function $G(\mathbf{k},i\omega_n;\mathbf{k}-2\mathbf{q},i\omega_n-2i\omega)$ can be expanded as a perturbation series of the unperturbed Green's function $G(\mathbf{k},i\omega_n)$.

For second-order response, the average value of the current Eq.~(\ref{3}) should have two vector potentials. If the current operator $\hat{j}_x$ takes the first term in Eq.~(\ref{2001}), the exact Green's function should contain two first term   or one second term  in Eq.~(\ref{2}). Similarly, if $\hat{j}_x$ takes the second term in Eq.~(\ref{2001}),  the exact Green's function should contain the first term in Eq.~(\ref{2}), and if $\hat{j}_x$ takes the third term in Eq.~(\ref{2001}), the exact Green's function does not contain any term of the interacting Hamiltonian.

For simplicity,
in this paper we consider a generic linear continuum Hamiltonian targeting various Dirac and Weyl systems.
Then, the corresponding Hamiltonian of the sample satisfies the restriction $\partial_{k_i} \partial_{k_j}H_0=\partial_{k_i} \hat{v}_j=0$.
As a result, the perturbed Hamiltonian becomes
\begin{equation}\label{200}
	H^{'}=e\mathbf{\hat{v}}\cdot \mathbf{A},
\end{equation}
the current operator is
\begin{equation}\label{201}
	\hat{j}_x=-e\partial_{k_x}H=-e\hat{v}_x,
\end{equation}
and the exact
Green's function is
given by the term
\begin{widetext}
\begin{eqnarray}\label{4} G(\mathbf{k},i\omega_n;\mathbf{k}-2\mathbf{q},i\omega_n-2i\omega)&=&2G(\mathbf{k},i\omega_n) H' (\mathbf{k},\mathbf{k}-\mathbf{q}) G(\mathbf{k}-\mathbf{q},i\omega_n-i\omega)H' (\mathbf{k}-\mathbf{q},\mathbf{k}-2\mathbf{q}) G(\mathbf{k}-2\mathbf{q},i\omega_n-2i\omega),\nonumber\\
\end{eqnarray}
\end{widetext}
where the unperturbed Green function has the form
\begin{equation}\label{5}
G(\mathbf{k},i\omega_n)=\sum_a \frac{\vert u_a(\mathbf{k})\rangle\langle u_a(\mathbf{k}) \vert}{i\omega_n +\mu-\varepsilon_a(\mathbf{k})}.
\end{equation}
 Here $\vert u_a(\mathbf{k})\rangle$ is the periodic part of Bloch wave functions for band $a$, $\mu$ is the chemical potential, and  $\varepsilon_a(\mathbf{k})$ is the energy dispersion of band $a$. 
The matrix element of the interaction Hamiltonian can be  obtained from 
\begin{widetext}
\begin{eqnarray}\label{55}
\langle u_a(\mathbf{k})|H^{'}(\mathbf{k},\mathbf{k}-\mathbf{q})|u_b(\mathbf{k}-\mathbf{q})\rangle&=&
\langle \psi_a(\mathbf{k})|H'|\psi_b(\mathbf{k}-\mathbf{q})\rangle\nonumber\\&=&\langle \psi_a(\mathbf{k})|\frac{e}{2}(\mathbf{\hat{v}}\cdot \mathbf{A}+\mathbf{A}\cdot \mathbf{\hat{v}})|\psi_b(\mathbf{k}-\mathbf{q})\rangle\nonumber\\&=&\langle u_a(\mathbf{k})|\frac{e}{2} A_y \big[ \hat{v}_y(\mathbf{k})+\hat{v}_y(\mathbf{k}-\mathbf{q})\big]|u_b(\mathbf{k}-\mathbf{q})\rangle,
\end{eqnarray}
\end{widetext}
where $|\psi_a(\mathbf{k})\rangle=e^{i\mathbf{k}\cdot \mathbf{r}}|u_a(\mathbf{k})\rangle$ is the Bloch wave function and $\hat{v}_y(\mathbf{k})=e^{-i\mathbf{k}\cdot \mathbf{r}}\hat{v}_ye^{i\mathbf{k}\cdot \mathbf{r}}$. Here we have symmetricized the perturbed Hamiltonian.
 For linear Hamiltonian, due to the fact that $\hat{v}_y(\mathbf{k}-\mathbf{q})=\hat{v}_y(\mathbf{k})-q\partial_{k_x}\hat{v}_y(\mathbf{k})=\hat{v}_y(\mathbf{k})$, we have \cite{araki2021spin,shi2007quantum,zhang2008theory}
 \begin{equation}\label{501}
H' (\mathbf{k},\mathbf{k}-\mathbf{q})=
\frac{e}{2}A_y \big[ \hat{v}_y(\mathbf{k})+\hat{v}_y(\mathbf{k}-\mathbf{q})\big]=eA_y \hat{v}_y(\mathbf{k}).
\end{equation}

\begin{figure}
	\centering
	\includegraphics[width=0.47\textwidth ]{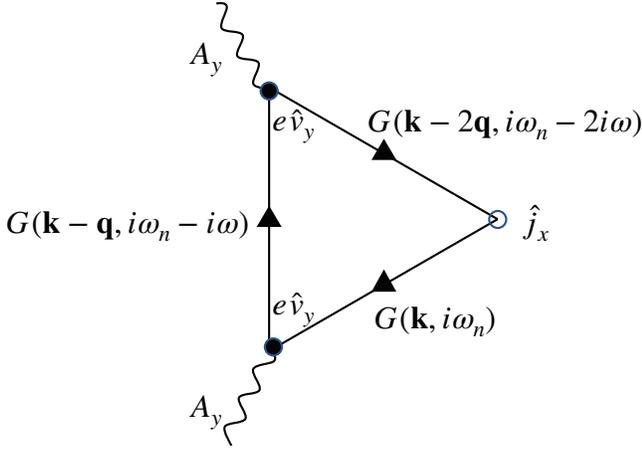}
	\caption{The triangle diagram of the second-order response current. The wavy lines refer to the external vector potential, the solid lines are the electron propagator, i.e., the Green's function,  the solid vertexes represent the operator $e\hat{v}_y$ in the perturbed Hamiltonian, and the hollow vertex denotes the Hall current operator.
	}\label{figure.2}
\end{figure}

 Note that we omit the time-dependent factor $e^{-i\omega t}$ in eq.~(\ref{55}), which  disappears in frequency space.
 The above second-order nonlinear response forms a triangle diagram, as shown in Fig.~\ref{figure.2}. 
 Besides, unlike the previous treatment of the second-order response \cite{gao2020second,parker2019diagrammatic}, here the wave vector $\mathbf{q}$ of the external fields is taken into account.

By taking the Matsubara sum over $i\omega_n$ in Eq.~(\ref{3}) and performing the analytical continuation $i\omega\rightarrow \omega +i0$, one obtains
\begin{equation}\label{6}
j_x(2\mathbf{q},2\omega)=\frac{-2e^3}{V}\Pi(\mathbf{q},\omega) A_y A_y,
\end{equation}
with 
\begin{equation}\label{7}
\Pi(\mathbf{q},\omega)=\sum_{\mathbf{k},a,b,c}M_{abc}(\mathbf{k},\mathbf{q})F_{abc}(\mathbf{k},\mathbf{q},\omega).
\end{equation}
Here $M_{abc}(\mathbf{k},\mathbf{q})$ is the correlation function with band indices $a$, $b$, and $c$:
\begin{widetext}
\begin{eqnarray}\label{8}
M_{abc}(\mathbf{k},\mathbf{q})&=&\langle u_c(\mathbf{k}-2\mathbf{q})\vert \hat{v}_x(\mathbf{k})\vert u_a(\mathbf{k})\rangle \langle u_a(\mathbf{k})\vert \hat{v}_y(\mathbf{k})\vert u_b(\mathbf{k}-\mathbf{q})\rangle\langle u_b(\mathbf{k}-\mathbf{q})\vert \hat{v}_y(\mathbf{k})\vert u_c(\mathbf{k}-2\mathbf{q})\rangle,
\end{eqnarray}
\end{widetext}
and
\begin{widetext}
\begin{eqnarray}\label{9}
F_{abc}(\mathbf{k},\mathbf{q},\omega)=\frac{1}{\varepsilon_a(\mathbf{k})-\varepsilon_b(\mathbf{k}-\mathbf{q})-\omega}\bigg[\frac{f(\varepsilon_a(\mathbf{k}))-f(\varepsilon_c(\mathbf{k}-2\mathbf{q}))}{\varepsilon_a(\mathbf{k})-\varepsilon_c(\mathbf{k}-2\mathbf{q})-2\omega}-\frac{f(\varepsilon_b(\mathbf{k}-\mathbf{q}))-f(\varepsilon_c(\mathbf{k}-2\mathbf{q}))}{\varepsilon_b(\mathbf{k}-\mathbf{q})-\varepsilon_c(\mathbf{k}-2\mathbf{q})-\omega}\bigg]
\end{eqnarray}
\end{widetext}
is the Matsubara summation with $f(\varepsilon_i(\mathbf{k}))=1/(e^{\beta(\varepsilon_i(\mathbf{k})-\mu)}+1)$ being the Fermi-Dirac distribution function.

\section{IV. Electromagnetic response} 
{Here we are interested in the dependence of the second harmonic response on the electric and magnetic field, i.e., $j_x(2\omega)  \propto E_y B_z $.
To derive the response to the electromagnetic fields, we need to expand  $\Pi(\mathbf{q},\omega)$ in Eq.~(\ref{6}) to first order in $q$ as
$\Pi(\mathbf{q},\omega)=\Pi(0,\omega)+q\Pi(\omega)$, where $\Pi(\omega)=\partial_{q} \Pi(\mathbf{q},\omega)\vert_{q=0}$.
As a result, Eq.~(\ref{6}) becomes
\begin{equation}\label{000}
j_x(2\omega)=\frac{2e^3}{\omega V}\Pi(\omega) E_y B_z.
\end{equation}
Note that the  $E^2$ and $B^2$ terms for the second harmonic generation can also be derived by expanding $\Pi(\mathbf{q},\omega)$ to the zero-order and second-order in  $q$, respectively.

In the low-frequency limit, the above current can be expanded in terms of the frequency $\omega$, i.e., $\Pi(\omega)/\omega=\Pi(0)/\omega+\partial_{\omega}\Pi(\omega)\vert_{\omega=0}+\dots$.
The second term has no dependence on $\omega$, it should generate an intrinsic nonlinear Hall current as shown in Ref. \cite{gao2014field}.
Here we are interested in the leading order term, i.e., 
\begin{equation}\label{0000}
j_x(2\omega)=\frac{2e^3}{\omega V}\Pi(0) E_y B_z.
\end{equation}
We note that the function $\Pi(\mathbf{q},\omega)$ satisfies the following relation (see Appendix B)
\begin{equation}\label{10}
\Pi(\mathbf{q},\omega)=\Pi^*(-\mathbf{q},-\omega).
\end{equation}
Then one can get
\begin{equation}\label{11}
\Pi(0)=-\Pi^*(0)=i \mathrm{lm} \Pi(0).
\end{equation}
The leading term of the current thus becomes
\begin{equation}\label{12}
j_x(2\omega)=\frac{2ie^3}{\omega V}\mathrm{lm}\Pi(0) E_y B_z.
\end{equation}

For two-band system, the energy bands $a,b,c$ in $M_{abc}(\mathbf{k},\mathbf{q})$ and $F_{abc}(\mathbf{k},\mathbf{q},\omega)$ can only be upper or lower bands. If the bands $a,b,c$ are the same band, $\mathrm{lm}M_{aaa}(\mathbf{k},0)$ and $F_{aaa}(\mathbf{k},0,\omega)$ will all be zero. Then  $\mathrm{lm}\Pi(\omega)=\partial_{q}\mathrm{lm}M_{aaa}(\mathbf{k},\mathbf{q})\vert_{q=0}F_{aaa}(\mathbf{k},0,\omega)+\mathrm{lm}M_{aaa}(\mathbf{k},0)\partial_{q}F_{aaa}(\mathbf{k},\mathbf{q},\omega)\vert_{q=0}=0$, i.e., the full intraband term's contribution to the second-order response vanishes. 
%
Note that for the linear response in the uniform limit, the intra-band terms also give zero contributions~\cite{chang2015chiral, zhong2016gyrotropic}. 
For nonzero response, the bands $a,b,c$ should take $a,b,b$; $a,b,a$; $a,a,b$.

One can decompose $\Pi(\mathbf{q},\omega)$ in Eq. (\ref{7}) into the intraband part $\Pi_\mathrm{intra}(\mathbf{q},\omega)$ and interband part 
$\Pi_\mathrm{inter}(\mathbf{q},\omega)$.
The intraband part $\Pi_\mathrm{intra}(\mathbf{q},\omega)$ is further split into three terms such that
$\Pi_\mathrm{intra}(\mathbf{q},\omega) = \sum_{\alpha=1}^3 \Pi^{(\alpha)}_\mathrm{intra}(\mathbf{q},\omega)$,
where
\begin{widetext}
\begin{equation}\label{s11}
\Pi^{(1)}_\mathrm{intra}(\mathbf{q},\omega)=\sum_{\mathbf{k},a\ne b}\frac{-M_{abb}(\mathbf{k},\mathbf{q})}{\varepsilon_a(\mathbf{k})-\varepsilon_b(\mathbf{k}-\mathbf{q})-\omega}\frac{f(\varepsilon_b(\mathbf{k}-\mathbf{q}))-f(\varepsilon_b(\mathbf{k}-2\mathbf{q}))}{\varepsilon_b(\mathbf{k}-\mathbf{q})-\varepsilon_b(\mathbf{k}-2\mathbf{q})-\omega},
\end{equation}
\begin{equation}\label{s12}
\Pi^{(2)}_\mathrm{intra}(\mathbf{q},\omega)=\sum_{\mathbf{k},a\ne b}\frac{M_{aba}(\mathbf{k},\mathbf{q})}{\varepsilon_a(\mathbf{k})-\varepsilon_b(\mathbf{k}-\mathbf{q})-\omega}\frac{f(\varepsilon_a(\mathbf{k}))-f(\varepsilon_a(\mathbf{k}-2\mathbf{q}))}{\varepsilon_a(\mathbf{k})-\varepsilon_a(\mathbf{k}-2\mathbf{q})-2\omega},
\end{equation}
and 
\begin{equation}\label{s13}
\Pi^{(3)}_\mathrm{intra}(\mathbf{q},\omega)=\sum_{\mathbf{k},a\ne b}\frac{M_{aab}(\mathbf{k},\mathbf{q})}{\varepsilon_a(\mathbf{k}-\mathbf{q})-\varepsilon_b(\mathbf{k}-2\mathbf{q})-\omega}\frac{f(\varepsilon_a(\mathbf{k}))-f(\varepsilon_a(\mathbf{k}-\mathbf{q}))}{\varepsilon_a(\mathbf{k})-\varepsilon_a(\mathbf{k}-\mathbf{q})-\omega}.
\end{equation}
\end{widetext}
From these terms, we expand the imaginary part of $\Pi_\mathrm{intra}(\mathbf{q},\omega)$, take the uniform limit, then obtain (see Appendix C)
\begin{equation}\label{19}
\mathrm{lm}\Pi_\mathrm{intra}(0)=-\sum_{\mathbf{k},a\ne b}F^{ab}_{xy}v_{ay}\partial_{k_x}f(\varepsilon_a(\mathbf{k})),
\end{equation}
where $F^{ab}_{xy}=-2\mathrm{lm}\langle \partial_{k_x}u_a(\mathbf{k})\vert  u_b(\mathbf{k})\rangle \langle u_b(\mathbf{k})\vert  \partial_{k_y} u_a(\mathbf{k})\rangle$ is the Berry curvature which is the imaginary part of quantum geometric tensor \cite{provost1980riemannian,zhang2022revealing} and $v_{iy}=\partial_{k_y}\varepsilon_i(\mathbf{k})$ is the band velocity along $y$-axis.

The interband part $\Pi_\mathrm{inter}(\mathbf{q},\omega)$ is also composed of three terms such that
$\Pi_\mathrm{inter}(\mathbf{q},\omega) = \sum_{\alpha=1}^3 \Pi^{(\alpha)}_\mathrm{inter}(\mathbf{q},\omega)$.
Since the interband part is irrespective of the order of limits $\mathbf{q}\rightarrow 0$ and $\omega\rightarrow 0$, we take $\omega= 0$ before $\mathbf{q}\rightarrow 0$ and have
\begin{widetext}
\begin{equation}\label{20}
    \Pi^{(1)}_\mathrm{inter}(\mathbf{q},0) =\sum_{\mathbf{k},a\ne b}\frac{M_{abb}(\mathbf{k},\mathbf{q})}{\varepsilon_a(\mathbf{k})-\varepsilon_b(\mathbf{k}-\mathbf{q})}\frac{f(\varepsilon_a(\mathbf{k}))-f(\varepsilon_b(\mathbf{k}-2\mathbf{q}))}{\varepsilon_a(\mathbf{k})-\varepsilon_b(\mathbf{k}-2\mathbf{q})},
\end{equation}
\begin{equation}\label{21}
    \Pi^{(2)}_\mathrm{inter}(\mathbf{q},0) = \sum_{\mathbf{k},a\ne b}\frac{-M_{aba}(\mathbf{k},\mathbf{q})}{\varepsilon_a(\mathbf{k})-\varepsilon_b(\mathbf{k}-\mathbf{q})}\frac{f(\varepsilon_a(\mathbf{k}-2\mathbf{q}))-f(\varepsilon_b(\mathbf{k}-\mathbf{q}))}{\varepsilon_a(\mathbf{k}-2\mathbf{q})-\varepsilon_b(\mathbf{k}-\mathbf{q})},
\end{equation}
and
\begin{equation}\label{22}
    \Pi^{(3)}_\mathrm{inter}(\mathbf{q},0) =\sum_{\mathbf{k},a\ne b}\frac{-M_{aab}(\mathbf{k},\mathbf{q})}{\varepsilon_a(\mathbf{k}-\mathbf{q})-\varepsilon_b(\mathbf{k}-2\mathbf{q})}\frac{f(\varepsilon_a(\mathbf{k}))-f(\varepsilon_b(\mathbf{k}-2\mathbf{q}))}{\varepsilon_a(\mathbf{k})-\varepsilon_b(\mathbf{k}-2\mathbf{q})}.
\end{equation}
\end{widetext}
We expand  $\mathrm{lm}\Pi_\mathrm{inter}(\mathbf{q},0)$ and get (see Appendix D)
\begin{equation}\label{23}
\mathrm{lm}\Pi_\mathrm{inter}(0)=\sum_{\mathbf{k},a\ne b}f(\varepsilon_a(\mathbf{k}))\partial_{k_x}\big[F^{ab}_{xy}(v_{ay}-v_{by})\big].
\end{equation}
Combining the contribution of the intraband and interband terms, the  Hall current becomes
\begin{equation}\label{24}
j_x(2\omega)=\frac{2ie^3}{\hbar\omega V}\sum_{\mathbf{k},a\ne b}f(\varepsilon_a(\mathbf{k}))\partial_{k_x}\big[F^{ab}_{xy}(2v_{ay}-v_{by})\big] E_y B_z.
\end{equation}
Here we have taken integration by parts for the Fermi surface term in Eq. (\ref{19}).
Eq. (\ref{24}) is the main result of this paper.
It shows that in the Hall device, there is a second-order Hall response which is proportional to a dipole, i.e.,
\begin{equation}\label{25}
D=\int[d\mathbf{k}]\sum_{a\ne b}f(\varepsilon_a(\mathbf{k}))\partial_{k_x}\big[F^{ab}_{xy}(2v_{ay}-v_{by})\big].
\end{equation}
Here $[d\mathbf{k}]=d^n \mathbf{k}/(2\pi)^n$ denotes the integration measure of a $n$-dimensional system. Under time-reversal symmetry, such a geometric quantity vanishes, since the partial differential, the Berry curvature, and the band velocity are all odd functions of momentum in such a case. The dipole can be written in another form with the term $\partial_{k_x}f(\epsilon_a(\mathbf{k}))$, which shows that its physical origin is the  Fermi surface.
 For simplicity, we only consider the clean limit
where the relaxation time $\tau$ approaches infinity.
Here the response depends on the electromagnetic field frequency $\omega$, which is similar to the extrinsic nonlinear Hall response \cite{sodemann2015quantum}. Note that in the clean limit, the factor $-\tau/(1+i\omega\tau)$
 in the extrinsic nonlinear Hall current is actually the factor $i/\omega$.

\begin{figure}[t]
\centering
\includegraphics[width=0.47\textwidth ]{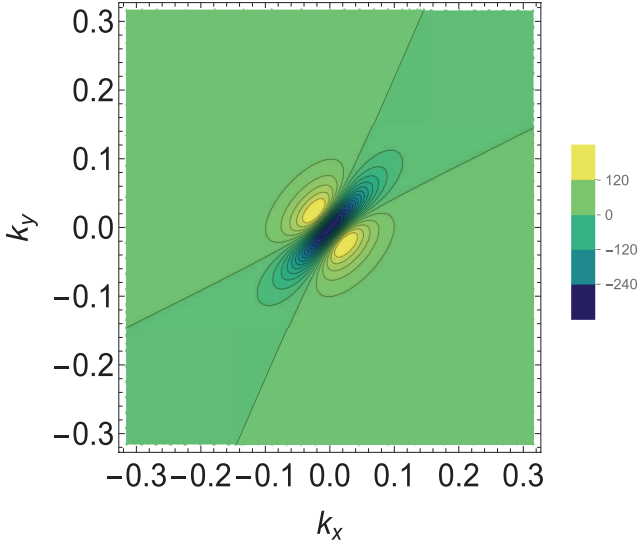}
\caption{The momentum space distribution of the dipole density.  Here the band gap $\Delta$ is chosen to be 0.1eV and the parameters $\alpha$, $\beta$ are set to be $\alpha/2=\beta$=1eV\AA.}
\label{figure.3}
\end{figure}

\section{V. Massive Dirac model} 
The result can be applied to Dirac and Weyl systems. Here, as an example, let us consider a two-dimensional Dirac Hamiltonian with broken time-reversal symmetry motivated by the LaAlO$_3$/LaNiO$_3$/LaAlO$_3$ quantum well system~\cite{tao2018two}.
The low-energy physics of the material around Dirac points can be described by the Hamiltonian
\begin{equation}\label{26}
H_0(\mathbf{k})=\alpha (k_x-k_y)\sigma_x+\beta (k_x+k_y)\sigma_z+\Delta \sigma_y.
\end{equation}
Here $\mathbf{k}$ is momentum defined near the Dirac point, $(\alpha, \beta)$ are expansion coefficients determined by the band parameters, and $\Delta$ is the gap at the Dirac points.
%
%
The energy dispersion of upper band $(+)$ and lower band $(-)$ are respectively given by $\varepsilon_{\pm}(\mathbf{k})=\pm \varepsilon_{0}$, where $\varepsilon_{0}=\sqrt{ k^{'2}_x+k^{'2}_y+\Delta^2 }$, $k^{'}_x=\alpha (k_x-k_y)$, and $k^{'}_y=\beta (k_x+k_y)$. 
Then the band velocity is $v_{\pm y}=\pm(-\alpha k^{'}_x+\beta k^{'}_y)/\varepsilon_{0}$ and
the Berry curvature is given by $F^{-+}_{xy}=-\alpha \beta \Delta /\varepsilon_{0}^{3}$. In the case of zero temperature and the chemical potential $\mu < -\Delta$, the dipole density, i.e., the integrand function in the dipole, becomes $3\alpha\beta\Delta [k^{'2}_x(3\alpha^2+\beta^2)-k^{'2}_y(\alpha^2+3\beta^2)+\Delta^2(\beta^2-\alpha^2)]/(4\pi^2\varepsilon_{0}^{6})$, $dk_xdk_y=dk^{'}_xdk^{'}_y/(2\alpha\beta)$.
and the integral is over the region $\varepsilon_{-}(\mathbf{k})< \mu $, i.e., $k^{'2}_x+k^{'2}_y>\mu^2-\Delta^2$. By using polar coordinate, the dipole is found to be
\begin{equation}\label{27}
D=\frac{3(\alpha^2-\beta^2)(\mu^2-\Delta^2)\Delta}{8\pi\hbar^2\mu^4}.
\end{equation}
%
In Figs.~\ref{figure.3}, we plot the distribution of the dipole density in the momentum space. It can be found that this quantity is concentrated in a small region in the momentum space and mirror symmetry $k_x\rightarrow-k_x$ and $k_y\rightarrow-k_y$ are broken.
We give the dipole dependence on the chemical potential $\mu$ in Fig.~\ref{figure.4}, which shows that the dipole is enhanced near the vertex of the band, i.e., $\mu=-\sqrt{2}\Delta$.

\begin{figure}[t]
\centering
\includegraphics[width=0.47\textwidth ]{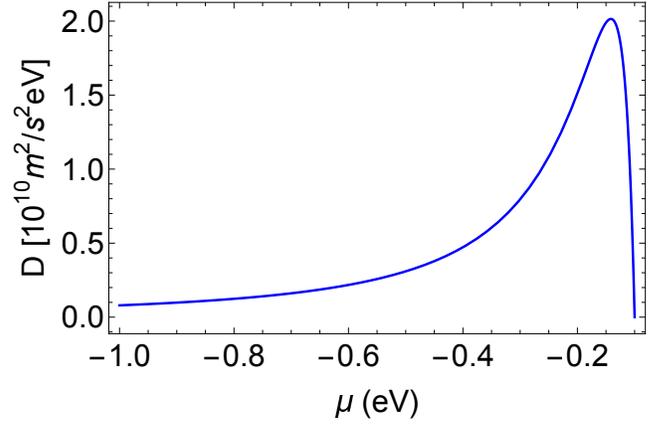}
\caption{The dipole as a function of chemical potential $\mu$. The parameters are the same as the setting in Figs.~\ref{figure.3}.
}\label{figure.4}
\end{figure}

\section{VI. Concluding remarks}
Let us estimate the magnitude of this nonlinear Hall response. Under a  driving magnetic field of $1$T and infrared photon energy $\hbar\omega=0.1$eV, the conductivity for this effect, i.e., $|j_x(2\omega)/E_y|=2e^3DB_z/\hbar\omega$, can reach $3.3\times 10^{-6}$A/V if we take $\alpha/2=\beta$=1eV\AA, $\Delta=2$ meV~\cite{tao2018two} and $\mu=-\sqrt{2}\Delta$. As a comparison, the magnitude of the conductivity for the quantum Hall effect is about $e^2/2\pi\hbar=3.9\times 10^{-5}$A/V. Thus this nonlinear Hall effect is large for low photon energy.

Here we mainly restrict our system to be linear. If there is a correction to the linear Hamiltonian, we should consider the second term in Eq.~(\ref{2}) and the last two terms in Eq.~(\ref{2001}). However, since these terms are given by Taylor series, all of them are small quantities. Thus it can be expected that the results in this paper will not alter qualitatively if there is a higher-order correction to the linear Hamiltonian.

In this work,  we consider the system where electric and magnetic fields are placed as shown in Fig.~\ref{figure.1}. Following a similar method, the results can be generalized to the system where electric and magnetic fields are placed in other ways.
Besides, by replacing the Hall current operator in the response with the spin current operator, the research can be extended to the spin system too \cite{sinova2004universal,zhang2022geometric}.


\section{Acknowledgments}
A.Z. acknowledges the support from Shanghai Magnolia Talent Plan  Youth Project and Shanghai Normal University (Grant No. 307-AF0102-24-005341).
 J.W.R was supported by the National Research Foundation of Korea (NRF) Grant funded by the Korean government (MSIT) (Grant Nos. 2021R1A2C1010572, 2021R1A5A1032996 and 2022M3H3A1063074) and the Ministry of Education (Grant No. RS-2023-00285390).
 



\section{Appendix A. The expectation value of the  current operator}

The expectation value of the  current operator can be expressed by  Green's function \cite{araki2021spin, vlasiuk2023cavity}. For readers to follow, here we give the  derivation.

By using the second quantization method, the expectation value of the  current operator can be written as
\begin{equation}\label{a01}
	\langle \hat{j_i}(\mathbf{r},\tau) \rangle=\sum_{m,n}j^{mn}_i\langle \hat{a}^\dag_{m}(\mathbf{r},\tau)  \hat{a}_{n}(\mathbf{r},\tau)\rangle.
\end{equation}
Here $\hat{a}^\dag_{m}(\mathbf{r},\tau),\hat{a}_{n}(\mathbf{r},\tau)$ are fermion creation and annihilation operators, respectively.
Note that using the Fourier transformation and Matsubara frequency, we have
\begin{equation}\label{a02}
	\hat{a}_{n}(\mathbf{r},\tau)=\frac{1}{(2\pi)^3\beta}\sum_{i\omega_n} \int d\mathbf{k} e^{i \mathbf{k} \cdot \mathbf{r}-i\omega_n\tau}\hat{a}_{n}(\mathbf{k},i\omega_n),
\end{equation}
where $\beta=1/k_B T$ is the inverse temperature and $\omega_n=(2n+1) \pi /\beta$ is the fermionic Matsubara frequency.
In momentum-frequency space, the expectation value of the second-order current  is
\begin{equation}\label{a03}
	\langle \hat{j_i}(2\mathbf{q},2i\omega) \rangle=\int d\mathbf{r}d\tau e^{-2i\mathbf{q}\cdot \mathbf{r}+2i\omega \tau}\langle \hat{j_i}(\mathbf{r},\tau) \rangle.
\end{equation}
Substituting Eq.~(\ref{a01})  into Eq.~(\ref{a03}) and using Eq.~(\ref{a02}), we obtain
\begin{eqnarray}\label{a04}
	\langle \hat{j_i}(2\mathbf{q},2i\omega) \rangle&=&\frac{1}{\beta V}\sum_{m,n,\mathbf{k},i\omega_n}j^{mn}_i\langle \hat{a}^\dag_{m}(\mathbf{k}-2\mathbf{q},i\omega_n-2i\omega)  \hat{a}_{n}(\mathbf{k},i\omega_n)\rangle\nonumber\\
	&=&\frac{1}{\beta V}\sum_{\mathbf{k},i\omega_n}\mathrm{Tr}[\hat{j_i}
	G(\mathbf{k},i\omega_n;\mathbf{k}-2\mathbf{q},i\omega_n-2i\omega)].
\end{eqnarray}
Here we have used the substitution $\frac{1}{2\pi}\int d\omega \leftrightarrow \frac{1}{\beta}\sum_{i\omega_n}$ and $\frac{1}{(2\pi)^3}\int d\mathbf{k} \leftrightarrow \frac{1}{V}\sum_{\mathbf{k}}$.

\section{Appendix B. The proof of the relation: $\Pi(\mathbf{q},\omega)=\Pi^*(-\mathbf{q},-\omega)$}

Here we show the details of the proof of the relation: $\Pi(\mathbf{q},\omega)=\Pi^*(-\mathbf{q},-\omega)$ in the main text, i.e., \begin{equation}\label{a1}
	\sum_{\mathbf{k},a,b,c}M_{abc}(\mathbf{k},\mathbf{q})F_{abc}(\mathbf{k},\mathbf{q},\omega)=\sum_{\mathbf{k},a,b,c}M^*_{abc}(\mathbf{k},-\mathbf{q})F_{abc}(\mathbf{k},-\mathbf{q},-\omega).
\end{equation}
We start from the expressions of $M_{abc}(\mathbf{k},\mathbf{q})$ and $F_{abc}(\mathbf{k},\mathbf{q},\omega)$, i.e.,
	\begin{eqnarray}\label{a2}
		M_{abc}(\mathbf{k},\mathbf{q})=&&\langle u_c(\mathbf{k}-2\mathbf{q})\vert \hat{v}_x\vert u_a(\mathbf{k})\rangle \langle u_a(\mathbf{k})\vert \hat{v}_y\vert u_b(\mathbf{k}-\mathbf{q})\rangle\nonumber\\&&
		\langle u_b(\mathbf{k}-\mathbf{q})\vert \hat{v}_y\vert u_c(\mathbf{k}-2\mathbf{q})\rangle
	\end{eqnarray}
and 
\begin{widetext}
	\begin{equation}\label{a3}
		F_{abc}(\mathbf{k},\mathbf{q},\omega)=\frac{1}{\varepsilon_a(\mathbf{k})-\varepsilon_b(\mathbf{k}-\mathbf{q})-\omega}\bigg[\frac{f(\varepsilon_a(\mathbf{k}))-f(\varepsilon_c(\mathbf{k}-2\mathbf{q}))}{\varepsilon_a(\mathbf{k})-\varepsilon_c(\mathbf{k}-2\mathbf{q})-2\omega}-\frac{f(\varepsilon_b(\mathbf{k}-\mathbf{q}))-f(\varepsilon_c(\mathbf{k}-2\mathbf{q}))}{\varepsilon_b(\mathbf{k}-\mathbf{q})-\varepsilon_c(\mathbf{k}-2\mathbf{q})-\omega}\bigg].
	\end{equation}
\end{widetext}
Then $M^*_{abc}(\mathbf{k},-\mathbf{q})$ is
\begin{widetext}
	\begin{equation}\label{a4}
		M^*_{abc}(\mathbf{k},-\mathbf{q})=\langle u_a(\mathbf{k})\vert \hat{v}_x\vert u_c(\mathbf{k}+2\mathbf{q})\rangle \langle u_c(\mathbf{k}+2\mathbf{q})\vert \hat{v}_y\vert u_b(\mathbf{k}+\mathbf{q})\rangle
		\langle u_b(\mathbf{k}+\mathbf{q})\vert \hat{v}_y\vert u_a(\mathbf{k})\rangle,
	\end{equation}
\end{widetext}
and $F^*_{abc}(\mathbf{k},-\mathbf{q},-\omega)$ becomes
\begin{widetext}
	\begin{equation}\label{a5}
		F_{abc}(\mathbf{k},-\mathbf{q},-\omega)=\frac{1}{\varepsilon_a(\mathbf{k})-\varepsilon_b(\mathbf{k}+\mathbf{q})+\omega}\bigg[\frac{f(\varepsilon_a(\mathbf{k}))-f(\varepsilon_c(\mathbf{k}+2\mathbf{q}))}{\varepsilon_a(\mathbf{k})-\varepsilon_c(\mathbf{k}+2\mathbf{q})+2\omega}-\frac{f(\varepsilon_b(\mathbf{k}+\mathbf{q}))-f(\varepsilon_c(\mathbf{k}+2\mathbf{q}))}{\varepsilon_b(\mathbf{k}+\mathbf{q})-\varepsilon_c(\mathbf{k}+2\mathbf{q})+\omega}\bigg].
	\end{equation}
\end{widetext}
Transferring the momentum from $\mathbf{k}$ to $\mathbf{k}-2\mathbf{q}$, $\sum_{\mathbf{k},a,b,c}M^*_{abc}(\mathbf{k},-\mathbf{q})F_{abc}(\mathbf{k},-\mathbf{q},-\omega)$ will be $\sum_{\mathbf{k},a,b,c}M^*_{abc}(\mathbf{k}-2\mathbf{q},-\mathbf{q})F_{abc}(\mathbf{k}-2\mathbf{q},-\mathbf{q},-\omega)$, where
\begin{widetext}
	\begin{equation}\label{a6}
		M^*_{abc}(\mathbf{k}-2\mathbf{q},-\mathbf{q})=\langle u_a(\mathbf{k}-2\mathbf{q})\vert \hat{v}_x\vert u_c(\mathbf{k})\rangle \langle u_c(\mathbf{k})\vert \hat{v}_y\vert u_b(\mathbf{k}-\mathbf{q})\rangle
		\langle u_b(\mathbf{k}-\mathbf{q})\vert \hat{v}_y\vert u_a(\mathbf{k}-2\mathbf{q})\rangle,
	\end{equation}
\end{widetext}
and
\begin{widetext}
	\begin{eqnarray}\label{a7}
		F_{abc}(\mathbf{k}-2\mathbf{q},-\mathbf{q},-\omega)=\frac{1}{\varepsilon_a(\mathbf{k}-2\mathbf{q})-\varepsilon_b(\mathbf{k}-\mathbf{q})+\omega}\bigg[\frac{f(\varepsilon_a(\mathbf{k}-2\mathbf{q}))-f(\varepsilon_c(\mathbf{k}))}{\varepsilon_a(\mathbf{k}-2\mathbf{q})-\varepsilon_c(\mathbf{k})+2\omega}-\frac{f(\varepsilon_b(\mathbf{k}-\mathbf{q}))-f(\varepsilon_c(\mathbf{k}))}{\varepsilon_b(\mathbf{k}-\mathbf{q})-\varepsilon_c(\mathbf{k})+\omega}\bigg].
	\end{eqnarray}
\end{widetext}
Now exchanging the indexes $a$ and $c$, i.e. $a\leftrightarrow c$, one gets $\sum_{\mathbf{k},a,b,c}M^*_{cba}(\mathbf{k}-2\mathbf{q},-\mathbf{q})F_{cba}(\mathbf{k}-2\mathbf{q},-\mathbf{q},-\omega)$, where
\begin{widetext}
	\begin{equation}\label{a8}
		M^*_{cba}(\mathbf{k}-2\mathbf{q},-\mathbf{q})=\langle u_c(\mathbf{k}-2\mathbf{q})\vert \hat{v}_x\vert u_a(\mathbf{k})\rangle \langle u_a(\mathbf{k})\vert \hat{v}_y\vert u_b(\mathbf{k}-\mathbf{q})\rangle
		\langle u_b(\mathbf{k}-\mathbf{q})\vert \hat{v}_y\vert u_c(\mathbf{k}-2\mathbf{q})\rangle=M_{abc}(\mathbf{q}),
	\end{equation}
\end{widetext}
and
\begin{widetext}
	\begin{eqnarray}\label{a9}
		F_{cba}(\mathbf{k}-2\mathbf{q},-\mathbf{q},-\omega)&=&\frac{1}{\varepsilon_c(\mathbf{k}-2\mathbf{q})-\varepsilon_b(\mathbf{k}-\mathbf{q})+\omega}\bigg[\frac{f(\varepsilon_c(\mathbf{k}-2\mathbf{q}))-f(\varepsilon_a(\mathbf{k}))}{\varepsilon_c(\mathbf{k}-2\mathbf{q})-\varepsilon_a(\mathbf{k})+2\omega}-\frac{f(\varepsilon_b(\mathbf{k}-\mathbf{q}))-f(\varepsilon_a(\mathbf{k}))}{\varepsilon_b(\mathbf{k}-\mathbf{q})-\varepsilon_a(\mathbf{k})+\omega}\bigg]\nonumber\\&=&F_{abc}(\mathbf{q},\omega).
	\end{eqnarray}
\end{widetext}
Thus we have
\begin{equation}\label{a10}
	\sum_{\mathbf{k},a,b,c}M^*_{abc}(-\mathbf{q})F_{abc}(-\mathbf{q},-\omega)=\sum_{\mathbf{k},a,b,c}M_{abc}(\mathbf{q})F_{abc}(\mathbf{q},\omega),
\end{equation}
i.e. the (\ref{a1}) is proved.

\section{Appendix C. The contribution of the intraband terms}

Here we show the contribution of the intraband terms in $\Pi(\mathbf{q},\omega)$.
The intraband terms in $\Pi(\mathbf{q},\omega)$ are respectively
\begin{widetext}
\begin{equation}\label{a11}
	\Pi^{(1)}_\mathrm{intra}(\mathbf{q},\omega)=\sum_{\mathbf{k},a\ne b}\frac{-M_{abb}(\mathbf{k},\mathbf{q})}{\varepsilon_a(\mathbf{k})-\varepsilon_b(\mathbf{k}-\mathbf{q})-\omega}\frac{f(\varepsilon_b(\mathbf{k}-\mathbf{q}))-f(\varepsilon_b(\mathbf{k}-2\mathbf{q}))}{\varepsilon_b(\mathbf{k}-\mathbf{q})-\varepsilon_b(\mathbf{k}-2\mathbf{q})-\omega},
\end{equation}
\begin{equation}\label{a12}
	\Pi^{(2)}_\mathrm{intra}(\mathbf{q},\omega)=\sum_{\mathbf{k},a\ne b}\frac{M_{aba}(\mathbf{k},\mathbf{q})}{\varepsilon_a(\mathbf{k})-\varepsilon_b(\mathbf{k}-\mathbf{q})-\omega}\frac{f(\varepsilon_a(\mathbf{k}))-f(\varepsilon_a(\mathbf{k}-2\mathbf{q}))}{\varepsilon_a(\mathbf{k})-\varepsilon_a(\mathbf{k}-2\mathbf{q})-2\omega},
\end{equation}
and 
\begin{equation}\label{a13}
	\Pi^{(3)}_\mathrm{intra}(\mathbf{q},\omega)=\sum_{\mathbf{k},a\ne b}\frac{M_{aab}(\mathbf{k},\mathbf{q})}{\varepsilon_a(\mathbf{k}-\mathbf{q})-\varepsilon_b(\mathbf{k}-2\mathbf{q})-\omega}\frac{f(\varepsilon_a(\mathbf{k}))-f(\varepsilon_a(\mathbf{k}-\mathbf{q}))}{\varepsilon_a(\mathbf{k})-\varepsilon_a(\mathbf{k}-\mathbf{q})-\omega}.
\end{equation}
	\end{widetext}
Expanding these intraband terms to first order in $q$, one can get
\begin{equation}\label{a14}
	\Pi^{(1)}_\mathrm{intra}(\mathbf{q},\omega)=q\sum_{\mathbf{k},a\ne b}\frac{M_{abb}(\mathbf{k},0)}{\varepsilon_a(\mathbf{k})-\varepsilon_b(\mathbf{k})-\omega}\frac{f'(\varepsilon_b(\mathbf{k}))}{\omega},
\end{equation}
\begin{equation}\label{a15}
	\Pi^{(2)}_\mathrm{intra}(\mathbf{q},\omega)=-q\sum_{\mathbf{k},a\ne b}\frac{M_{aba}(\mathbf{k},0)}{\varepsilon_a(\mathbf{k})-\varepsilon_b(\mathbf{k})-\omega}\frac{f'(\varepsilon_a(\mathbf{k}))}{\omega},
\end{equation}
and 
\begin{equation}\label{a16}
	\Pi^{(3)}_\mathrm{intra}(\mathbf{q},\omega)=-q\sum_{\mathbf{k},a\ne b}\frac{M_{aab}(\mathbf{k},0)}{\varepsilon_a(\mathbf{k})-\varepsilon_b(\mathbf{k})-\omega}\frac{f'(\varepsilon_a(\mathbf{k}))}{\omega}.
\end{equation}
Here $f'=\partial_{k_x}f$. We note that $M_{aba}(\mathbf{k},0)=\langle u_a(\mathbf{k})\vert \hat{v}_x(\mathbf{k})\vert u_a(\mathbf{k})\rangle \langle u_a(\mathbf{k})\vert \hat{v}_y(\mathbf{k})\vert u_b(\mathbf{k})\rangle\langle u_b(\mathbf{k})\vert \hat{v}_y(\mathbf{k})\vert u_a(\mathbf{k})\rangle$ which is a real number, so $\Pi^{(2)}_\mathrm{intra}(\mathbf{q},\omega)$ can be omitted since it does not contribute to the response. Besides, after exchanging the indexes $a, b$ for $\Pi^{(1)}_\mathrm{intra}(\mathbf{q},\omega)$, one has
\begin{eqnarray}\label{a17}
	\Pi^{(1)}_\mathrm{intra}(\mathbf{q},\omega)&=&-q\sum_{\mathbf{k},a\ne b}\frac{M_{baa}(\mathbf{k},0)}{\varepsilon_a(\mathbf{k})-\varepsilon_b(\mathbf{k})+\omega}\frac{f'(\varepsilon_a(\mathbf{k}))}{\omega}\nonumber\\&=&q\sum_{\mathbf{k},a\ne b}\frac{M_{baa}(\mathbf{k},0)f'(\varepsilon_a(\mathbf{k}))}{\varepsilon_a(\mathbf{k})-\varepsilon_b(\mathbf{k})}\bigg(\frac{1}{\varepsilon_a(\mathbf{k})-\varepsilon_b(\mathbf{k})}-\frac{1}{\omega}\bigg),\nonumber\\
\end{eqnarray}
where
\begin{eqnarray}\label{a18}
	M_{baa}(\mathbf{k},0)&=&\langle u_a(\mathbf{k})\vert \hat{v}_x(\mathbf{k})\vert u_b(\mathbf{k})\rangle \langle u_b(\mathbf{k})\vert \hat{v}_y(\mathbf{k})\vert u_a(\mathbf{k})\rangle\nonumber\\&&\langle u_a(\mathbf{k})\vert \hat{v}_y(\mathbf{k})\vert u_a(\mathbf{k})\rangle,
\end{eqnarray}
and we have expanded $\Pi^{(1)}_\mathrm{intra}(\mathbf{q},\omega)$ in the limit $\omega\rightarrow 0$.
Note that $M_{aab}(\mathbf{k},0)$ in $\Pi^{(3)}_\mathrm{intra}(\mathbf{q},\omega)$ is
\begin{eqnarray}\label{a19}
	M_{aab}(\mathbf{k},0)&=&\langle u_b(\mathbf{k})\vert \hat{v}_x(\mathbf{k})\vert u_a(\mathbf{k})\rangle \langle u_a(\mathbf{k})\vert \hat{v}_y(\mathbf{k})\vert u_a(\mathbf{k})\rangle\nonumber\\&&\langle u_a(\mathbf{k})\vert \hat{v}_y(\mathbf{k})\vert u_b(\mathbf{k})\rangle
	\nonumber\\&=&M^*_{baa}(\mathbf{k},0).
\end{eqnarray}
Thus
\begin{equation}\label{170}
	\Pi^{(3)}_\mathrm{intra}(\mathbf{q},\omega)=-q\sum_{\mathbf{k},a\ne b}\frac{M^*_{baa}(\mathbf{k},0)f'(\varepsilon_a(\mathbf{k}))}{\varepsilon_a(\mathbf{k})-\varepsilon_b(\mathbf{k})}\bigg(\frac{1}{\varepsilon_a(\mathbf{k})-\varepsilon_b(\mathbf{k})}+\frac{1}{\omega}\bigg),
\end{equation}
and then
\begin{eqnarray}\label{171}
	\mathrm{lm}\Pi_\mathrm{intra}(\mathbf{q},\omega)&=&2q\sum_{\mathbf{k},a\ne b}\frac{\mathrm{lm}M_{baa}(\mathbf{k},0)f'(\varepsilon_a(\mathbf{k}))}{(\varepsilon_a(\mathbf{k})-\varepsilon_b(\mathbf{k}))^2}	\nonumber\\&=&-q\sum_{\mathbf{k},a\ne b}F^{ab}_{xy}v_{ay}\partial_{k_x}f(\varepsilon_a(\mathbf{k})),
\end{eqnarray}
i.e., the intraband terms in $\Pi(\mathbf{q},\omega)$ contributes $-\sum_{\mathbf{k},a\ne b}F^{ab}_{xy}v_{ay}\partial_{k_x}f(\varepsilon_a(\mathbf{k}))$ to $\mathrm{lm}\Pi(0)$.
Here $F^{ab}_{xy}=-2\mathrm{lm}\langle u_a(\mathbf{k})\vert \hat{v}_x(\mathbf{k})\vert u_b(\mathbf{k})\rangle \langle u_b(\mathbf{k})\vert \hat{v}_y(\mathbf{k})\vert u_a(\mathbf{k})\rangle/(\varepsilon_a(\mathbf{k})-\varepsilon_b(\mathbf{k}))^2$ is the Berry curvature.

\section{Appendix D. The contribution of the interband terms}

Here we show the derivation of Eq.~(\ref{23}) in the main text.
The interband terms in $\Pi(\mathbf{q},\omega)$ can be respectively decomposed  as
\begin{eqnarray}\label{a20}
	&&\sum_{\mathbf{k},a\ne b}\frac{M_{abb}(\mathbf{k},\mathbf{q})}{\varepsilon_a(\mathbf{k})-\varepsilon_b(\mathbf{k}-\mathbf{q})}\frac{f(\varepsilon_a(\mathbf{k}))-f(\varepsilon_b(\mathbf{k}-2\mathbf{q}))}{\varepsilon_a(\mathbf{k})-\varepsilon_b(\mathbf{k}-2\mathbf{q})}\nonumber\\&&=\sum_{\mathbf{k},a\ne b}\frac{M_{abb}(\mathbf{k},\mathbf{q})}{\varepsilon_a(\mathbf{k})-\varepsilon_b(\mathbf{k}-\mathbf{q})}\frac{f(\varepsilon_a(\mathbf{k}))}{\varepsilon_a(\mathbf{k})-\varepsilon_b(\mathbf{k}-2\mathbf{q})}\nonumber\\
	&&-\sum_{\mathbf{k},a\ne b}\frac{M_{baa}(\mathbf{k},\mathbf{q})}{\varepsilon_a(\mathbf{k}-\mathbf{q})-\varepsilon_b(\mathbf{k})}\frac{f(\varepsilon_a(\mathbf{k}-2\mathbf{q}))}{\varepsilon_a(\mathbf{k}-2\mathbf{q})-\varepsilon_b(\mathbf{k})},
\end{eqnarray}
\begin{eqnarray}\label{a21}
	&&\sum_{\mathbf{k},a\ne b}\frac{-M_{aba}(\mathbf{k},\mathbf{q})}{\varepsilon_a(\mathbf{k})-\varepsilon_b(\mathbf{k}-\mathbf{q})}\frac{f(\varepsilon_a(\mathbf{k}-2\mathbf{q}))-f(\varepsilon_b(\mathbf{k}-\mathbf{q}))}{\varepsilon_a(\mathbf{k}-2\mathbf{q})-\varepsilon_b(\mathbf{k}-\mathbf{q})}\nonumber\\&&=\sum_{\mathbf{k},a\ne b}\frac{-M_{aba}(\mathbf{k},\mathbf{q})}{\varepsilon_a(\mathbf{k})-\varepsilon_b(\mathbf{k}-\mathbf{q})}\frac{f(\varepsilon_a(\mathbf{k}-2\mathbf{q}))}{\varepsilon_a(\mathbf{k}-2\mathbf{q})-\varepsilon_b(\mathbf{k}-\mathbf{q})}\nonumber\\
	&&+\sum_{\mathbf{k},a\ne b}\frac{M_{bab}(\mathbf{k},\mathbf{q})}{\varepsilon_a(\mathbf{k}-\mathbf{q})-\varepsilon_b(\mathbf{k})}\frac{f(\varepsilon_a(\mathbf{k}-\mathbf{q}))}{\varepsilon_a(\mathbf{k}-\mathbf{q})-\varepsilon_b(\mathbf{k}-2\mathbf{q})},
\end{eqnarray}
and
\begin{eqnarray}\label{a22}
	&&\sum_{\mathbf{k},a\ne b}\frac{-M_{aab}(\mathbf{k},\mathbf{q})}{\varepsilon_a(\mathbf{k}-\mathbf{q})-\varepsilon_b(\mathbf{k}-2\mathbf{q})}\frac{f(\varepsilon_a(\mathbf{k}))-f(\varepsilon_b(\mathbf{k}-2\mathbf{q}))}{\varepsilon_a(\mathbf{k})-\varepsilon_b(\mathbf{k}-2\mathbf{q})}\nonumber\\&&=\sum_{\mathbf{k},a\ne b}\frac{-M_{aab}(\mathbf{k},\mathbf{q})}{\varepsilon_a(\mathbf{k}-\mathbf{q})-\varepsilon_b(\mathbf{k}-2\mathbf{q})}\frac{f(\varepsilon_a(\mathbf{k}))}{\varepsilon_a(\mathbf{k})-\varepsilon_b(\mathbf{k}-2\mathbf{q})}\nonumber\\
	&&+\sum_{\mathbf{k},a\ne b}\frac{M_{bba}(\mathbf{k},\mathbf{q})}{\varepsilon_a(\mathbf{k}-2\mathbf{q})-\varepsilon_b(\mathbf{k}-\mathbf{q})}\frac{f(\varepsilon_a(\mathbf{k}-2\mathbf{q}))}{\varepsilon_a(\mathbf{k}-2\mathbf{q})-\varepsilon_b(\mathbf{k})}.
\end{eqnarray}

We note that $M_{abb}(\mathbf{k},\mathbf{q})=\langle u_b(\mathbf{k}-2\mathbf{q})\vert \hat{v}_x(\mathbf{k})\vert u_a(\mathbf{k})\rangle \langle u_a(\mathbf{k})\vert \hat{v}_y(\mathbf{k})\vert u_b(\mathbf{k}-\mathbf{q})\rangle \langle u_b(\mathbf{k}-\mathbf{q})\vert \hat{v}_y(\mathbf{k})\vert u_b(\mathbf{k}-2\mathbf{q})\rangle$ can be expanded as
\begin{equation}\label{a23}
	M_{abb}(\mathbf{k},\mathbf{q})=M_{abb}(\mathbf{k},0)+q\partial_q M_{abb}(\mathbf{k},\mathbf{q})\vert_{q=0},
\end{equation}
where $M_{abb}(\mathbf{k},0)= \langle u_a(\mathbf{k})\vert \hat{v}_y(\mathbf{k})\vert u_b(\mathbf{k})\rangle \langle u_b(\mathbf{k})\vert \hat{v}_x(\mathbf{k})\vert u_a(\mathbf{k})\rangle v_{by}$,
and $\partial_q M_{abb}(\mathbf{k},\mathbf{q})\vert_{q=0}$ is
\begin{eqnarray}\label{a24}
&&-\partial_{k_x}M_{abb}(\mathbf{k},0)-\langle \partial_{k_x}u_b(\mathbf{k})\vert u_a(\mathbf{k})\rangle  \langle u_a(\mathbf{k})\vert \hat{v}_y(\mathbf{k})\vert u_b(\mathbf{k})\rangle v_{ax}v_{by}\nonumber\\
	&&-\langle u_b(\mathbf{k})\vert \hat{v}_x(\mathbf{k})\vert u_a(\mathbf{k})\rangle \langle u_a(\mathbf{k})\vert \hat{v}_y(\mathbf{k})\vert u_b(\mathbf{k})\rangle\langle u_b(\mathbf{k})\vert \hat{v}_y(\mathbf{k})\vert u_a(\mathbf{k})\rangle\nonumber\\
	&&\cdot\langle u_a(\mathbf{k})\vert \partial_{k_x}u_b(\mathbf{k})\rangle+\langle u_b(\mathbf{k})\vert \partial_{k_x} u_a(\mathbf{k})\rangle \langle u_a(\mathbf{k})\vert \hat{v}_y(\mathbf{k})\vert u_b(\mathbf{k})\rangle v_{bx}v_{by}\nonumber\\&&+\langle u_b(\mathbf{k})\vert \hat{v}_x(\mathbf{k})\vert u_a(\mathbf{k})\rangle \langle \partial_{k_x}u_a(\mathbf{k})\vert u_b(\mathbf{k})\rangle v_{by}v_{by}\nonumber\\&=&-\partial_{k_x}M_{abb}(\mathbf{k},0)-\langle \partial_{k_x}u_a(\mathbf{k})\vert  u_b(\mathbf{k})\rangle\langle u_b(\mathbf{k})\vert \hat{v}_y(\mathbf{k})\vert u_a(\mathbf{k})\rangle \nonumber\\&&\cdot(v_{ax}+v_{bx})v_{by}.
\end{eqnarray}
Here $v_{ai}=\partial_{k_i}\varepsilon_a(\mathbf{k})$ and we have omitted real terms.
Besides,
\begin{eqnarray}\label{a25}
	&&\frac{1}{\varepsilon_a(\mathbf{k})-\varepsilon_b(\mathbf{k}-\mathbf{q})}\frac{f(\varepsilon_a(\mathbf{k}))}{\varepsilon_a(\mathbf{k})-\varepsilon_b(\mathbf{k}-2\mathbf{q})}\nonumber\\&=&\bigg[\frac{1}{(\varepsilon_a(\mathbf{k})-\varepsilon_b(\mathbf{k}))^2}-3q\frac{v_{bx}}{(\varepsilon_a(\mathbf{k})-\varepsilon_b(\mathbf{k}))^3}\bigg]f(\varepsilon_a(\mathbf{k})).\nonumber\\
\end{eqnarray}
So the first term of Eq. (\ref{a20}) contributes 
\begin{equation}\label{a26}
	\sum_{\mathbf{k},a\ne b}\mathrm{lm}\bigg[\frac{\partial_q M_{abb}(\mathbf{k},\mathbf{q})\vert_{q=0}}{(\varepsilon_a(\mathbf{k})-\varepsilon_b(\mathbf{k}))^2}-3\frac{M_{abb}(\mathbf{k},0)v_{bx}}{(\varepsilon_a(\mathbf{k})-\varepsilon_b(\mathbf{k}))^3}\bigg]f(\varepsilon_a(\mathbf{k})
\end{equation}
to the $\mathrm{lm}\Pi(0)$ in the current.

The second term of  Eq. (\ref{a20}) can be rewritten as
\begin{eqnarray}\label{a27}
	&&-\sum_{\mathbf{k},a\ne b}\frac{M_{baa}(\mathbf{k},\mathbf{q})}{\varepsilon_a(\mathbf{k}-\mathbf{q})-\varepsilon_b(\mathbf{k})}\frac{f(\varepsilon_a(\mathbf{k}-2\mathbf{q}))}{\varepsilon_a(\mathbf{k}-2\mathbf{q})-\varepsilon_b(\mathbf{k})}\nonumber\\&=&-\sum_{\mathbf{k},a\ne b}\frac{M_{baa}(\mathbf{k}+2\mathbf{q},\mathbf{q})}{\varepsilon_a(\mathbf{k}+\mathbf{q})-\varepsilon_b(\mathbf{k}+2\mathbf{q})}\frac{f(\varepsilon_a(\mathbf{k}))}{\varepsilon_a(\mathbf{k})-\varepsilon_b(\mathbf{k}+2\mathbf{q})}.\nonumber\\
\end{eqnarray}
Here $M_{baa}(\mathbf{k}+2\mathbf{q},\mathbf{q})=\langle u_a(\mathbf{k})\vert \hat{v}_x(\mathbf{k})\vert u_b(\mathbf{k}+2\mathbf{q})\rangle \langle u_b(\mathbf{k}+2\mathbf{q})\vert \hat{v}_y(\mathbf{k})\vert u_a(\mathbf{k}+\mathbf{q})\rangle \langle u_a(\mathbf{k}+\mathbf{q})\vert \hat{v}_y(\mathbf{k})\vert u_a(\mathbf{k})\rangle$ can be decomposed as
\begin{eqnarray}\label{a28}
	M_{baa}(\mathbf{k}+2\mathbf{q},\mathbf{q})&=&M_{baa}(\mathbf{k},0)+q\partial_q M_{baa}(\mathbf{k}+2\mathbf{q},\mathbf{q})\vert_{q=0}\nonumber\\
\end{eqnarray}
with
$M_{baa}(\mathbf{k},0)=\langle u_a(\mathbf{k})\vert \hat{v}_x(\mathbf{k})\vert u_b(\mathbf{k})\rangle \langle u_b(\mathbf{k})\vert \hat{v}_y(\mathbf{k})\vert u_a(\mathbf{k})\rangle v_{ay}$ and
\begin{eqnarray}\label{a29}
	\partial_q M_{baa}(\mathbf{k}+2\mathbf{q},\mathbf{q})\vert_{q=0}&=&
	\partial_{k_x}M_{baa}(\mathbf{k},0)-\langle \partial_{k_x} u_a(\mathbf{k})\vert u_b(\mathbf{k})\rangle \nonumber\\&&\cdot\langle u_b(\mathbf{k}) \hat{v}_y(\mathbf{k})\vert u_a(\mathbf{k})\rangle (v_{ax}+v_{bx})v_{ay}.\nonumber\\
\end{eqnarray}
Besides,
\begin{widetext}
\begin{equation}\label{a30}
	\frac{-1}{\varepsilon_a(\mathbf{k}+\mathbf{q})-\varepsilon_b(\mathbf{k}+2\mathbf{q})}\frac{f(\varepsilon_a(\mathbf{k}))}{\varepsilon_a(\mathbf{k})-\varepsilon_b(\mathbf{k}+2\mathbf{q})}=\bigg[\frac{-1}{(\varepsilon_a(\mathbf{k})-\varepsilon_b(\mathbf{k}))^2}+q\frac{v_{ax}-4v_{bx}}{(\varepsilon_a(\mathbf{k})-\varepsilon_b(\mathbf{k}))^3}\bigg]f(\varepsilon_a(\mathbf{k})).
\end{equation}
\end{widetext}
So the second term of Eq. (\ref{a20}) contributes 
\begin{widetext}
\begin{equation}\label{a31}
	\sum_{\mathbf{k},a\ne b}\mathrm{lm}\bigg[\frac{-\partial_q M_{baa}(\mathbf{k}+2\mathbf{q},\mathbf{q})\vert_{q=0}}{(\varepsilon_a(\mathbf{k})-\varepsilon_b(\mathbf{k}))^2}+\frac{M_{baa}(\mathbf{k},0)(v_{ax}-4v_{bx})}{(\varepsilon_a(\mathbf{k})-\varepsilon_b(\mathbf{k}))^3}\bigg]f(\varepsilon_a(\mathbf{k}))
\end{equation}
\end{widetext}
to the $\mathrm{lm}\Pi(0)$ in the current.

Similarly, for the terms in Eqs. (\ref{a21}) and (\ref{a22}), the contribution to the $\mathrm{lm}\Pi(0)$ are respectively
\begin{widetext}
\begin{equation}\label{a32}
	\sum_{\mathbf{k},a\ne b}\mathrm{lm}\bigg[\frac{-\partial_q M_{aba}(\mathbf{k}+2\mathbf{q},\mathbf{q})\vert_{q=0}}{(\varepsilon_a(\mathbf{k})-\varepsilon_b(\mathbf{k}))^2}+2\frac{M_{aba}(\mathbf{k},0)(v_{ax}-v_{bx})}{(\varepsilon_a(\mathbf{k})-\varepsilon_b(\mathbf{k}))^3}\bigg]f(\varepsilon_a(\mathbf{k})),
\end{equation}
\end{widetext}
\begin{equation}\label{a33}
	\sum_{\mathbf{k},a\ne b}\mathrm{lm}\bigg[\frac{\partial_q M_{bab}(\mathbf{k}+\mathbf{q},\mathbf{q})\vert_{q=0}}{(\varepsilon_a(\mathbf{k})-\varepsilon_b(\mathbf{k}))^2}\bigg]f(\varepsilon_a(\mathbf{k})),
\end{equation}
\begin{equation}\label{a34}
	\sum_{\mathbf{k},a\ne b}\mathrm{lm}\bigg[\frac{-\partial_q M_{aab}(\mathbf{k},\mathbf{q})\vert_{q=0}}{(\varepsilon_a(\mathbf{k})-\varepsilon_b(\mathbf{k}))^2}-\frac{M_{aab}(\mathbf{k},0)(v_{ax}-4v_{bx})}{(\varepsilon_a(\mathbf{k})-\varepsilon_b(\mathbf{k}))^3}\bigg]f(\varepsilon_a(\mathbf{k})),
\end{equation}
and 
\begin{equation}\label{a35}
	\sum_{\mathbf{k},a\ne b}\mathrm{lm}\bigg[\frac{\partial_q M_{bba}(\mathbf{k}+2\mathbf{q},\mathbf{q})\vert_{q=0}}{(\varepsilon_a(\mathbf{k})-\varepsilon_b(\mathbf{k}))^2}+3\frac{M_{bba}(\mathbf{k},0)v_{bx}}{(\varepsilon_a(\mathbf{k})-\varepsilon_b(\mathbf{k}))^3}\bigg]f(\varepsilon_a(\mathbf{k})),
\end{equation}
where
\begin{widetext}
\begin{equation}\label{a36}
	M_{aba}(\mathbf{k},0)=\langle u_a(\mathbf{k})\vert \hat{v}_y(\mathbf{k})\vert u_b(\mathbf{k})\rangle\langle u_b(\mathbf{k})\vert \hat{v}_y(\mathbf{k})\vert u_a(\mathbf{k})\rangle v_{ax}---real, discarded,
\end{equation}
\begin{equation}\label{a37}
	\partial_q M_{aba}(\mathbf{k}+2\mathbf{q},\mathbf{q})\vert_{q=0}=2\langle \partial_{k_x}u_a(\mathbf{k})\vert  u_b(\mathbf{k})\rangle\langle u_b(\mathbf{k})\vert \hat{v}_y(\mathbf{k})\vert u_a(\mathbf{k})\rangle v_{ax}v_{by},
\end{equation}
\begin{equation}\label{a38}
	\partial_q M_{bab}(\mathbf{k}+\mathbf{q},\mathbf{q})\vert_{q=0}=2\langle \partial_{k_x}u_a(\mathbf{k})\vert  u_b(\mathbf{k})\rangle\langle u_b(\mathbf{k})\vert \hat{v}_y(\mathbf{k})\vert u_a(\mathbf{k})\rangle v_{bx}v_{ay},
\end{equation}
\begin{equation}\label{a39}
	M_{aab}(\mathbf{k},0)= \langle u_a(\mathbf{k})\vert \hat{v}_y(\mathbf{k})\vert u_b(\mathbf{k})\rangle\langle u_b(\mathbf{k})\vert \hat{v}_x(\mathbf{k})\vert u_a(\mathbf{k})\rangle v_{ay},
\end{equation}
\begin{equation}\label{a40}
	\partial_q M_{aab}(\mathbf{k},\mathbf{q})\vert_{q=0}=-\partial_{k_x}M_{aab}(\mathbf{k},0)-\langle \partial_{k_x}u_a(\mathbf{k})\vert  u_b(\mathbf{k})\rangle\langle u_b(\mathbf{k})\vert \hat{v}_y(\mathbf{k})\vert u_a(\mathbf{k})\rangle (v_{ax}+v_{bx})v_{ay},
\end{equation}
\begin{equation}\label{a41}
	M_{bba}(\mathbf{k},0)=\langle u_a(\mathbf{k})\vert \hat{v}_x(\mathbf{k})\vert u_b(\mathbf{k})\rangle \langle u_b(\mathbf{k})\vert \hat{v}_y(\mathbf{k})\vert u_a(\mathbf{k})\rangle v_{by},
\end{equation}
\begin{equation}\label{a42}
	\partial_q M_{bba}(\mathbf{k}+2\mathbf{q},\mathbf{q})\vert_{q=0}=\partial_{k_x}M_{bba}(\mathbf{k},0)-\langle \partial_{k_x}u_a(\mathbf{k})\vert  u_b(\mathbf{k})\rangle\langle u_b(\mathbf{k})\vert \hat{v}_y(\mathbf{k})\vert u_a(\mathbf{k})\rangle (v_{ax}+v_{bx})v_{by}.
\end{equation}
\end{widetext}

Sum the terms in Eqs. (\ref{a26}), (\ref{a31}), (\ref{a32}), (\ref{a33}), (\ref{a34}), and (\ref{a35}) up, one obtains
\begin{widetext}
\begin{eqnarray}\label{a43}
	\mathrm{lm}\Pi(0)&=&\sum_{\mathbf{k},a\ne b}f(\varepsilon_a(\mathbf{k}))\bigg\{\frac{2\partial_{k_x}\big[\mathrm{lm}\langle u_a(\mathbf{k})\vert \hat{v}_x(\mathbf{k})\vert u_b(\mathbf{k})\rangle \langle u_b(\mathbf{k})\vert \hat{v}_y(\mathbf{k})\vert u_a(\mathbf{k})\rangle (v_{by}-v_{ay})\big]}{(\varepsilon_a(\mathbf{k})-\varepsilon_b(\mathbf{k}))^2}\nonumber\\&&+\frac{4\mathrm{lm}\langle u_a(\mathbf{k})\vert \hat{v}_x(\mathbf{k})\vert u_b(\mathbf{k})\rangle \langle u_b(\mathbf{k})\vert \hat{v}_y(\mathbf{k})\vert u_a(\mathbf{k})\rangle\nonumber(v_{ax}-v_{bx})(v_{ay}-v_{by})}{(\varepsilon_a(\mathbf{k})-\varepsilon_b(\mathbf{k}))^3}\bigg\}\\&=&
	\sum_{\mathbf{k},a\ne b}f(\varepsilon_a(\mathbf{k}))\partial_{k_x}\big[F^{ab}_{xy}(v_{ay}-v_{by})\big].
\end{eqnarray}
\end{widetext}


\begin{thebibliography}{81}
\bibitem{gao2014field} Y. Gao, S. A. Yang, and Q. Niu, Field induced positional shift of Bloch electrons and its dynamical implications, Phys. Rev. Lett. $\textbf{112}$, 166601 (2014).
\bibitem{liu2021intrinsic}
H. Liu, J. Zhao, Y. -X. Huang, W. Wu, X. -L. 
Sheng, C. Xiao,  and S. A. Yang, Intrinsic Second-Order Anomalous Hall Effect and Its Application in Compensated Antiferromagnets, Phys. Rev. Lett. {\bf127}, 277202 (2021).
\bibitem{wang2021intrinsic}
C. Wang, Y. Gao,  and D. Xiao, Intrinsic nonlinear Hall effect in antiferromagnetic tetragonal CuMnAs, Phys. Rev.
Lett. {\bf127}, 277201 (2021).
\bibitem{bharti2022high}
A. Bharti, M. Mrudul, and G. Dixit, High-harmonic spectroscopy of light-driven nonlinear anisotropic anomalous Hall effect in a Weyl semimetal, Phys. Rev. B {\bf105}, 155140 (2022).
\bibitem{bhalla2022resonant}
P. Bhalla, K. Das, D. Culcer, and A. Agarwal, Resonant second-harmonic generation as a probe of quantum geometry, Phys. Rev. Lett. {\bf129}, 227401 (2022).
\bibitem{sodemann2015quantum}
 I. Sodemann and L. Fu,  Quantum nonlinear Hall effect induced by Berry curvature dipole in time-reversal invariant materials, Phys. Rev. Lett. {\bf115}, 216806 (2015).
\bibitem{morimoto2016semiclassical}
T. Morimoto, S. Zhong, J. Orenstein, and J. E. Moore, Semiclassical theory of nonlinear magneto-optical responses with applications to topological Dirac/Weyl semimetals, 
Phys. Rev. B {\bf94}, 245121 (2016).
\bibitem{facio2018strongly}
J. I. Facio, D. Efremov, K. Koepernik, J. -S. You, I. Sodemann, and J. Van Den Brink, Strongly enhanced Berry dipole at topological phase transitions in BiTeI, Phys. Rev. Lett. {\bf121}, 246403 (2018).
\bibitem{you2018berry} J. -S. You, S. Fang, S.-Y. Xu, E. Kaxiras, and T. Low, Berry curvature dipole current in the transition metal dichalcogenides family, Phys. Rev. B $\textbf{98}$, 121109 (2018).
 \bibitem{zhang2018berry} Y. Zhang, Y. Sun, and B. Yan, Berry curvature dipole in Weyl semimetal materials: an ab initio study, Phys. Rev. B $\textbf{97}$, 041101 (2018).
\bibitem{ma2019observation} Q. Ma S.-Y. Xu, H. Shen, D. MacNeill, V. Fatemi, T. -R. Chang, A. M. Mier Valdivia, S. Wu, Z. Du, C. -H. Hsu, et al, Observation of the nonlinear Hall effect under time-reversal-symmetric conditions, Nature {\bf565}, 337 (2019).
\bibitem{kang2019nonlinear} K. Kang, T. Li, E. Sohn, J. Shan, and K. F. Mak, Nonlinear anomalous Hall effect in few-layer WTe2, Nat. Mater. $\textbf{18}$, 324 (2019).
\bibitem{gao2020second}  Y. Gao, F. Zhang, and W. Zhang,  Second-order nonlinear Hall effect in Weyl semimetals, Phys. Rev. B {\bf102}, 245116 (2020).
\bibitem{watanabe2021chiral} H. Watanabe and Y. Yanase, Chiral photocurrent in parity-violating magnet and enhanced response in topological antiferromagnet, Phys. Rev. X $\textbf{11}$, 011001 (2021).
\bibitem{du2021quantum}  Z. Du, C. Wang, H. -P. Sun, H. -Z. Lu, and X. Xie,  Quantum theory of the nonlinear Hall effect, Nat. Commun. {\bf12}, 1 (2021).
\bibitem{du2021nonlinear} Z. Du, H. -Z. Lu, and X. Xie, Nonlinear hall effects, Nat. Rev. Phys. {\bf3}, 744 (2021).
\bibitem{wang2024orbital} L. Wang, J. Zhu, H. Chen, H. Wang, J. Liu, Y.-X. Huang,
B. Jiang, J. Zhao, H. Shi, G. Tian, et al., Orbital magneto-nonlinear anomalous Hall effect in kagome magnet Fe3Sn2, Phys. Rev. Lett. {\bf132}, 106601 (2024).
\bibitem{shao2020nonlinear} D.-F. Shao, S.-H. Zhang, G. Gurung, W. Yang, and E. Y.
Tsymbal, Nonlinear anomalous Hall effect for N{\'e}el vector detection, Phys.  Rev. Lett. {\bf124}, 067203 (2020).
\bibitem{rostami2020probing} H. Rostami and V. Juri{\v{c}}i{\'c},  Probing quantum criticality using nonlinear Hall effect in a metallic Dirac system, Phys. Rev. Res. {\bf2}, 013069 (2020).
\bibitem{araki2021spin}  Y. Araki, D. Suenaga, K. Suzuki, and S. Yasui,  Spin-orbital magnetic response of relativistic fermions with band hybridization, Phys. Rev. Res. {\bf3}, 023098 (2021).
\bibitem{shi2007quantum}  J. Shi, G. Vignale, D. Xiao, and Q. Niu, Quantum theory of orbital magnetization and its generalization to interacting systems, Phys. Rev. Lett. {\bf99}, 197202 (2007).
\bibitem{zhang2008theory} P. Zhang, Z. Wang, J. Shi, D. Xiao, and Q. Niu, Theory of conserved spin current and its application to a two-dimensional hole gas, Phys. Rev. B {\bf77}, 075304 (2008).
\bibitem{parker2019diagrammatic}  D. E. Parker, T. Morimoto, J. Orenstein, and J. E.
Moore, Diagrammatic approach to nonlinear optical response with application to Weyl semimetals, Phys. Rev. B {\bf99}, 045121 (2019).
\bibitem{chang2015chiral} M.-C. Chang and M.-F. Yang, Chiral magnetic effect in a two-band lattice model of Weyl semimetal, Phys. Rev. B {\bf91}, 115203 (2015).
\bibitem{zhong2016gyrotropic} S. Zhong, J. E. Moore, and I. Souza, Gyrotropic magnetic effect and the magnetic moment on the Fermi surface, Phys. Rev. Lett. {\bf116}, 077201 (2016).
\bibitem{provost1980riemannian} J. Provost and G. Vallee,  Riemannian structure on manifolds of quantum states, Commun. Math. Phys. {\bf76}, 289 (1980).
\bibitem{zhang2022revealing} A. Zhang, Revealing Chern number from quantum metric, Chin. Phys.  B {\bf31}, 040201 (2022).
\bibitem{tao2018two}  L. Tao and E. Y. Tsymbal,  Two-dimensional type-II Dirac fermions in a LaAlO3/LaNiO3/LaAlO3 quantum well, Phys. Rev. B {\bf98}, 121102 (2018).
\bibitem{sinova2004universal} J. Sinova, D. Culcer, Q. Niu, N. Sinitsyn, T. Jungwirth,
and A. H. MacDonald, Universal intrinsic spin Hall effect, Phys. Rev. Lett. {\bf92}, 126603 (2004).
\bibitem{zhang2022geometric}  A. Zhang and J.-W. Rhim,  Geometric origin of intrinsic spin hall effect in an inhomogeneous electric field, Commun. Phys. {\bf5}, 195 (2022).
\bibitem{vlasiuk2023cavity} E. Vlasiuk, V.K. Kozin, J. Klinovaja, D. Loss, I. V.  Iorsh, and I. V. Tokatly,
 Cavity-induced charge transfer in periodic systems: Length-gauge formalism, Phys. Rev. B {\bf108}, 085410 (2023).
\end{thebibliography}

\end{document}